\documentstyle[amssymb,aps]{revtex}

\begin{document}
\title{Adiabatic creation of entangled states by a bichromatic field\\
designed from the topology of the dressed eigenenergies}
\author{S. Gu\'{e}rin$^{1}$, R. Unanyan $^{2,3}$, L. P. Yatsenko$^{4}$ and
H. R. Jauslin$^{1}$}
\address{$^{1}$Laboratoire de Physique de l'Universit\'{e} de Bourgogne,
CNRS, 
BP 47870, 21078 Dijon, France\\
$^{2}$Fachbereich Physik, Universit\"at Kaiserslautern, 67653 Kaiserslautern,%
Germany\\
$^{3}$Institute for Physical Research, Armenian National Academy of
Sciences, 378410 Ashtarak, Armenia\\
$^{4}$Institute of Physics of the Ukrainian Academy of Sciences, prospekt
Nauky, 46, 252650 Kiev-22, Ukraine}
\maketitle

\begin{abstract}
Preparation of entangled pairs of coupled two-state systems driven by a
bichromatic external field is studied. We use a system of two coupled spin-$%
\frac{1}{2}$ that can be translated into a three-state ladder model whose
intermediate state represents the entangled state. We show that this
entangled state can be prepared in a robust way with appropriate fields.
Their frequencies and envelopes are derived from the topological properties
of the model.
\end{abstract}


Email: sguerin@u-bourgogne.fr

\section{Introduction}

Entanglement is a key concept in various contemporary areas of active
research in quantum physics. It explicitly demonstrates the non-local
character of quantum theory, having potential applications in quantum
communication, cryptography and computation \cite{Williams}. The preparation
of an entangled state is of great interest for both fundamental and applied
reasons. During the last few years various methods for preparation of
entangled states of atomic systems have been proposed and some of them
experimentally demonstrated \cite{Walther,Molmer}.

Although a quantum system can be manipulated by tailored sequences of
resonant pulses of precise area, in particular $\pi $ and $\pi /2$ pulses
respectively for the complete inversion and the equal weight coherent
superposition, deviations from the precise pulse area and from resonance can
lead to significant errors. Adiabatic passage techniques provide much
greater robustness against fluctuations in the interaction parameters. The
stimulated Raman adiabatic passage (STIRAP)\ method \cite{Bergmann} has been
proposed for the creation of an entangled state of two three-level atoms in
a QED cavity \cite{Pellizzari} and for $\Lambda $ atomic systems \cite%
{Zadkov}.

In this paper we propose a simple method for entangling two subsystems
driven by pulse shaped external fields. For definiteness we take these to be
two identical spins interacting with each other and driven by
radio-frequency fields. This system can be translated in a three-level
ladder model with the intermediate level corresponding to the entangled
state \cite{UnanyanRMN}. The goal is to populate completely this entangled
state at the end of the pulses. The most efficient couplings are obtained
with two near one-photon resonant fields. We show that unlike in the STIRAP
process, one- and two-photon detunings are required to populate most
efficiently the intermediate level.

\section{The model: Two spin system in external fields}

We consider two spin-$\frac{1}{2}$ particles coupled by uniaxial exchange
interaction in the $z$ direction \cite{UnanyanRMN}. In a time-dependent
magnetic field ${\bf B}(t)=[B_{x}(t),B_{y}(t),B_{z}]$, the Hamiltonian of
this system reads ($\hbar =1$) 
\begin{equation}
{\sf \hat{H}}(t)={\sf \hat{H}}_{0}+\mu {\bf B}(t)\cdot \left( {\bf \hat{S}}%
_{1}+{\bf \hat{S}}_{2}\right) ,  \label{Hamiltonian}
\end{equation}
where 
\begin{equation}
{\sf \hat{H}}_{0}=4\xi \hat{S}_{1}^{z}\otimes \hat{S}_{2}^{z}  \label{H0}
\end{equation}
is the part describing the exchange interaction, $\mu $ is the gyromagnetic
ratio, $\xi $ is the exchange-interaction constant, and ${\bf \hat{S}}_{k}=[%
\hat{S}_{k}^{x},\hat{S}_{k}^{y},\hat{S}_{k}^{z}]$ is the $k$th spin operator 
$(k=1,2)$. We assume that the static magnetic $B_{z}$ field in the $z$
direction is strong enough ($|\mu B_{z}|\gg |\xi |$) so that the model with
exchange interaction only in the $z$ direction is justified.

In the spin product state space $\{\left| m\right\rangle _{1}\left|
m\right\rangle _{2}\}$ $(m=\downarrow ,\uparrow )$, where the states $\left|
\downarrow \right\rangle _{k}$ and $\left| \uparrow \right\rangle _{k}$
denote respectively the spin-down and spin-up states of the $k$th spin, a
complete basis of orthonormalized eigenstates of $\hat{H}_{0}$ is given by 
\begin{mathletters}
\begin{eqnarray}
\left| \downarrow \downarrow \right\rangle &\equiv &\left| \downarrow
\right\rangle _{1}\left| \downarrow \right\rangle _{2},  \label{Phi00} \\
\left| \downarrow \uparrow ^{+}\right\rangle &\equiv &\frac{1}{\sqrt{2}}%
\left[ \left| \downarrow \right\rangle _{1}\left| \uparrow \right\rangle
_{2}+\left| \uparrow \right\rangle _{1}\left| \downarrow \right\rangle _{2}%
\right] ,  \label{Phi12plus} \\
\left| \uparrow \uparrow \right\rangle &\equiv &\left| \uparrow
\right\rangle _{1}\left| \uparrow \right\rangle _{2},  \label{Phi11} \\
\left| \downarrow \uparrow ^{-}\right\rangle &\equiv &\frac{1}{\sqrt{2}}%
\left[ \left| \downarrow \right\rangle _{1}\left| \uparrow \right\rangle
_{2}-\left| \uparrow \right\rangle _{1}\left| \downarrow \right\rangle _{2}%
\right] .  \label{Phi12minus}
\end{eqnarray}
In this basis, the Hamiltonian (\ref{Hamiltonian}) can be expressed in the
block-matrix form 
\end{mathletters}
\begin{equation}
{\sf H}(t)=\left[ 
\begin{array}{cc}
{\bf H}_{c}(t) & {\bf 0} \\ 
{\bf 0} & -\xi%
\end{array}
\right] ,  \label{H}
\end{equation}
where 
\begin{equation}
{\bf H}_{\text{c}}=\left[ 
\begin{array}{ccc}
\xi -\beta _{z} & \frac{1}{\sqrt{2}}(\beta _{x}+i\beta _{y}) & 0 \\ 
\frac{1}{\sqrt{2}}(\beta _{x}-i\beta _{y}) & -\xi & \frac{1}{\sqrt{2}}(\beta
_{x}+i\beta _{y}) \\ 
0 & \frac{1}{\sqrt{2}}(\beta _{x}-i\beta _{y}) & \xi +\beta _{z}%
\end{array}
\right]  \label{Hc}
\end{equation}
with ${\bf \beta \equiv \lbrack }\beta _{x},\beta _{y},\beta _{z}]=\mu {\bf B%
}$. The state $\left| \downarrow \uparrow ^{-}\right\rangle $ is thus
decoupled from the other states; it describes the evolution of a spin$-0$
singlet in a time-dependent magnetic field. This decoupling justifies our
choice of the basis. The three other states $\left| \downarrow \downarrow
\right\rangle $, $\left| \downarrow \uparrow ^{+}\right\rangle $ and $\left|
\uparrow \uparrow \right\rangle $ are coupled by the transverse $(xy)$
magnetic field. To complete the definition of the problem, we suppose that
initially the two-spin system is in the unentangled state $\left| \downarrow
\downarrow \right\rangle $. Our goal is to establish the conditions leading
to the most efficient robust transfer into the entangled state $\left|
\downarrow \uparrow ^{+}\right\rangle $.

We consider the case when the spin system interacts with a constant magnetic
field in the $z$ direction and two radio-frequency fields of respective
frequencies $\omega _{1}$ and $\omega _{2}$ in the $x$ direction: 
\begin{mathletters}
\label{beta}
\begin{eqnarray}
\beta _{z} &=&\text{const}, \\
\beta _{x} &=&\Omega _{1}\left( t\right) \cos \left( \omega _{1}t+\varphi
_{1}\right) +\Omega _{2}\left( t\right) \cos \left( \omega _{2}t+\varphi
_{2}\right) , \\
\beta _{y} &=&0.
\end{eqnarray}
The state vector $\phi (t)$ is solution of the Schr\"{o}dinger equation i$%
\frac{d}{dt}\phi (t)={\sf H}_{\text{c}}(t)\phi (t)$ with the Hamiltonian $%
{\sf H}_{\text{c}}(t)$ (\ref{Hc}) written in the basis $\left\{ \left|
\downarrow \downarrow \right\rangle ,\left| \downarrow \uparrow
^{+}\right\rangle ,\left| \uparrow \uparrow \right\rangle \right\} $. When
the radiofrequency fields are off ($\beta _{x}=0$), we have the following
energies $E_{\downarrow \downarrow }\equiv \xi -\beta _{z}$, $E_{\downarrow
\uparrow ^{+}}\equiv -\xi $ and $E_{\uparrow \uparrow }\equiv \xi +\beta
_{z} $. Without loss of generality we assume $\xi <0$ and $\beta _{z}>0,$
leading for $\beta _{z}>2\left| \xi \right| $ to a ladder configuration $%
E_{\downarrow \downarrow }<E_{\downarrow \uparrow ^{+}}<E_{\uparrow \uparrow
}$. We apply near resonant fields $\omega _{1}\approx E_{\downarrow \uparrow
^{+}}-E_{\downarrow \downarrow }$, $\omega _{2}\approx E_{\uparrow \uparrow
}-E_{\downarrow \uparrow ^{+}}$, i.e. with the detunings
$\Delta _{1}$ and $\Delta _{2}$: 
\end{mathletters}
\begin{mathletters}
\label{omega}
\begin{eqnarray}
\omega _{1} &\equiv &2\left| \xi \right| +\beta _{z}-\Delta _{1}, \\
\omega _{2} &\equiv &-2\left| \xi \right| +\beta _{z}-\Delta _{2}.
\end{eqnarray}
According to the rotating wave approximation (RWA), one can neglect non
resonant counterrotating terms under the conditions $\omega _{1,2}\gg \Omega
_{1}\left( t\right) ,\Omega _{2}\left( t\right) $. The RWA transformation 
\end{mathletters}
\begin{equation}
{\sf R}=\left[ 
\begin{array}{ccc}
e^{-\text{i}E_{\downarrow \downarrow }t} & 0 & 0 \\ 
0 & e^{-\text{i}\left( E_{\downarrow \downarrow }+\omega _{1}\right) t} & 0
\\ 
0 & 0 & e^{-\text{i}\left( E_{\downarrow \downarrow }+\omega _{1}+\omega
_{2}\right) t}%
\end{array}
\right]
\end{equation}
leads to the state vector $\widetilde{\phi }(t)={\sf R}^{\dagger }\phi (t)$
(whose coefficients have the same absolute values as the ones of $\phi $),
that satisfies the Schr\"{o}dinger equation 
\begin{equation}
\text{i}\frac{d}{dt}\widetilde{\phi }(t)=\widetilde{{\sf H}}_{\text{c}}(t)%
\widetilde{\phi }(t),  \label{SE}
\end{equation}
with the Hamiltonian $\widetilde{{\sf H}}_{\text{c}}={\sf R}^{\dagger }{\sf H%
}_{\text{c}}{\sf R}-$i$\left( \partial {\sf R}^{\dagger }/\partial t\right) 
{\sf R}$, where only the quasi-resonant terms have been kept: 
\begin{equation}
\widetilde{{\sf H}}_{\text{c}}=\frac{1}{2}\left[ 
\begin{array}{ccc}
0 & \Omega _{1}\left( t\right) +e^{-\text{i}\delta t}\Omega _{2}\left(
t\right) & 0 \\ 
\Omega _{1}\left( t\right) +e^{\text{i}\delta t}\Omega _{2}\left( t\right) & 
2\Delta _{1} & \Omega _{2}\left( t\right) +e^{\text{i}\delta t}\Omega
_{1}\left( t\right) \\ 
0 & \Omega _{2}\left( t\right) +e^{-\text{i}\delta t}\Omega _{1}\left(
t\right) & 2\left( \Delta _{1}+\Delta _{2}\right)%
\end{array}
\right] .  \label{Heff}
\end{equation}
The frequency 
\begin{equation}
\delta \equiv \omega _{1}-\omega _{2}=4\left| \xi \right| +\Delta
_{2}-\Delta _{1}  \label{defdelta}
\end{equation}
characterizes the {\it coupling ambiguity} \cite{confusion}. This
Hamiltonian $\widetilde{{\sf H}}_{\text{c}}$ allows indeed both fields to
couple the two transitions when the field amplitudes $\Omega _{1}\left(
t\right) $ and $\Omega _{2}\left( t\right) $ are not small compared to $%
\delta $. The competing coupling schemes are depicted in Fig. \ref{linkage}
with $\left| 1\right\rangle \equiv \left| \downarrow \downarrow
\right\rangle ,$ $\left| 2\right\rangle \equiv \left| \downarrow \uparrow
^{+}\right\rangle $ and $\left| 3\right\rangle \equiv \left| \uparrow
\uparrow \right\rangle $. Two limit channels can thus be exhibited: the
channel A (respectively B) corresponding to the absorptions of one $\omega
_{1}-$photon (respectively of one $\omega _{2}-$photon) by the $1-2$
transition and of one $\omega _{2}-$photon (respectively of one $\omega
_{1}- $photon) by the $2-3$ transition. We will study more precisely in
Sections V and VI the various regimes that occur in this system. The problem
of preparing the entangled state $\left| 2\right\rangle \equiv \left|
\downarrow \uparrow ^{+}\right\rangle $ is thus reduced to the study of the
population transfer of the intermediate level in the ladder system driven by
the Hamiltonian $\widetilde{{\sf H}}_{\text{c}}$ (\ref{Heff}).

The populations given by the Schr\"{o}dinger equation (\ref{SE}) are
invariant under the following transformation ${\cal T}$%
\begin{mathletters}
\label{symmetry}
\begin{eqnarray}
\Delta _{1} &\rightarrow &\Delta _{1}+\delta , \\
\Delta _{2} &\rightarrow &\Delta _{2}-\delta , \\
\delta &\rightarrow &-\delta , \\
\Omega _{1} &\rightleftharpoons &\Omega _{2}.
\end{eqnarray}
We indeed obtain $\widetilde{{\sf R}}^{\dagger }\left( {\cal T}\widetilde{%
{\sf H}}_{\text{c}}\right) \widetilde{{\sf R}}-$i$\left( \partial \widetilde{%
{\sf R}}^{\dagger }/\partial t\right) \widetilde{{\sf R}}=\widetilde{{\sf H}}%
_{\text{c}}$, with the unitary transformation 
\end{mathletters}
\begin{equation}
\widetilde{{\sf R}}=\left[ 
\begin{array}{ccc}
1 & 0 & 0 \\ 
0 & e^{-\text{i}\delta t} & 0 \\ 
0 & 0 & 1%
\end{array}
\right] .
\end{equation}

\section{Numerical results}

Figures \ref{contours1} and \ref{contours2} display the population of the
state $\left| 2\right\rangle $ at the end of a sequence of delayed gaussian
pulses of the same lengths and the same peak amplitudes 
\begin{mathletters}
\label{gauss}
\begin{eqnarray}
\Omega _{1}\left( t\right) &=&\Omega _{0}\exp \left[ -\left( t+\tau \right)
^{2}/T^{2}\right] , \\
\Omega _{2}\left( t\right) &=&\Omega _{0}\exp \left[ -\left( t-\tau \right)
^{2}/T^{2}\right] ,
\end{eqnarray}
for various normalized peak amplitudes $\Omega _{0}/\delta $ and detunings $%
\Delta /\delta $, where we have chosen 
\end{mathletters}
\begin{equation}
\Delta \equiv \Delta _{1}=\Delta _{2}.
\end{equation}
We have considered this restriction of the parameters because it gives
preferentially large islands of good population transfer. This will be
justified in Sections V and VI. Note that the case $\Delta _{1}=-\Delta _{2}$
is irrelevant since it corresponds to a two-photon resonance between the
product states $\left| \downarrow\downarrow \right\rangle$ and $\left|
\uparrow\uparrow \right\rangle$. We could have considered equivalently the
restriction $\Delta _{2}=\Delta _{1}+2\delta $ in accordance with the
symmetry (\ref{symmetry}). The two different orderings of pulses have been
considered: the sequence 1 of Fig. \ref{contours1} (respectively the
sequence 2 of Fig. \ref{contours2}) corresponds to the $\omega _{1}-$pulse
(respectively the $\omega _{2}-$pulse) being switched on first, with the
delay $\tau =1.7T$ (respectively $\tau =-1.7T$). Global adiabaticity is
ensured by the choice of a large pulse area $\Omega _{0}T=50$.

One can distinguish three islands of robust high transfer (white
regions). Specific parameters characterizing these islands are labelled by ($%
a$), ($a^{\prime }$), ($d$), and ($d^{\prime }$), with the subscript 1 or 2
respectively for the figures \ref{contours1} or \ref{contours2} [except ($%
a_{2}$) which is outside the regions of high transfer]. These islands of
high transfer are analyzed in the following Sections by using the dressed
Hamiltonian corresponding to $\widetilde{{\sf H}}_{\text{c}}$ (\ref{Heff})
and the adiabatic properties of the dynamics. We will characterize different
regimes and associate them with different effective dressed Hamiltonians. We
will show that the islands of good transfer can be understood from the
topological properties of the appropriate effective dressed Hamiltonian.

We will show the following results:

(i) Regions (a) correspond to a STIRAP-like process associated to the
channel A (see Fig. \ref{linkage}) that is perturbed (in the sense of
non-resonant perturbation theory) by the channel B. Note that the
restriction $\Delta _{2}=\Delta _{1}+2\delta $ would have given a
STIRAP-like process associated to the channel B perturbed by the channel A;

(ii) Regions (d)\ (in the weak field regime, i.e. $\Omega _{1},\Omega
_{2}<\delta $) correspond to an effective two-level SCRAP-like (Stark
chirped rapid adiabatic passage) process \cite{Yatsenko_PRA99,Rickes};

(iii) Regions (d')\ (in the strong field regime, i.e. $\Omega _{1},\Omega
_{2}\gtrsim \delta $) corresponds to an effective two-level bichromatic
SCRAP process (with additional Stark shifts) as described in Ref. \cite%
{PRA_RC}.

\section{The dressed Hamiltonian}

It is convenient to use the adiabatic Floquet theory in order to study the
Hamiltonian $\widetilde{{\sf H}}_{\text{c}}$ (\ref{Heff}) since its time
dependence contains a characteristic frequency $\delta $. The dressed
Hamiltonian (or quasienergy Hamiltonian) corresponding to $\widetilde{{\sf H}%
}_{\text{c}}$ is \cite{confusion,OE99}: 
\begin{equation}
{\sf K}^{\left[ \Omega _{1},\Omega _{2}\right] }=-\text{i}\delta \frac{%
\partial }{\partial \theta }+\frac{1}{2}\left[ 
\begin{array}{ccc}
0 & \Omega _{1}\left( t\right) +e^{-\text{i}\theta }\Omega _{2}\left(
t\right) & 0 \\ 
\Omega _{1}\left( t\right) +e^{\text{i}\theta }\Omega _{2}\left( t\right) & 
2\Delta _{1} & \Omega _{2}\left( t\right) +e^{\text{i}\theta }\Omega
_{1}\left( t\right) \\ 
0 & \Omega _{2}\left( t\right) +e^{-\text{i}\theta }\Omega _{1}\left(
t\right) & 2\left( \Delta _{1}+\Delta _{2}\right)%
\end{array}
\right] .  \label{K}
\end{equation}
This dressed Hamiltonian depends parametrically on the pulse shapes and the
detunings. It acts in the Hibert space spanned on the three states $\left\{
\left| 1\right\rangle ,\left| 2\right\rangle ,\left| 3\right\rangle \right\} 
$ tensored by the photonic Hilbert space \cite{PRA_RC,confusion}. The
eigenstates of ${\sf K}$ are families of three states denoted $\left|
1;k,-k\right\rangle $, $\left| 2;k-1,-k\right\rangle $ and $\left|
3;k-1,-k-1\right\rangle $ with $k$ a positive or negative integer. The
corresponding eigenvalues $\lambda _{1;k,-k}$, $\lambda _{2;k-1,-k}$ and $%
\lambda _{3;k-1,-k-1}$ have the following periodicity property: $\lambda
_{n;k_{1},k_{2}}=\lambda _{n;k_{1}-1,k_{2}+1}+\hbar \delta $, for $n=1,2,3$.
The notation $\left| n;k_{1},k_{2}\right\rangle $ characterizes (when the
fields are off) the state $\left| n\right\rangle $ dressed by the field of $%
k_{1}$ $\omega _{1}-$photons and of $k_{2}$ $\omega _{2}-$photons. The
integers $k_{1}$ and $k_{2}$ characterize thus relative photon numbers of
the respective fields of frequency $\omega _{1}$ and $\omega _{2}.$ The
initial state is denoted $\left| 1;0,0\right\rangle .$ The problem can be
formulated as follows: {\it we look for robust adiabatic connections between
the initial state }$\left| 1;0,0\right\rangle ${\it \ and the final state }$%
\left| 2;k-1,-k\right\rangle ${\it \ for some positive or negative integer }$%
k$.

The possible connections depend on the {\it topology} of the eigenenergy
surfaces of (\ref{K}) as functions of the field envelopes $\Omega _{1}$ and $%
\Omega _{2}$ for given $\Delta _{1}$ and $\Delta _{2}$ \cite{PRA_RC,topology}%
. The topology is characterized by true crossings which occur generically
when one of the fields is off. We will study in the following the topology
of ${\sf K}$ using different effective dressed Hamiltonians corresponding to
different regimes. These regimes will depend on the ranges of the detunings
and of the field amplitudes.

We classify the different regimes and construct effective dressed
Hamiltonians by determining in the Hamilonian ${\sf K}$ (\ref{K}) which
terms are {\it resonant} (or {\it quasi-resonant}) and which one are only 
{\it perturbative}. The resonant terms are treated by an adapted unitary
transformation which allows an explicit diagonalization, whereas the
perturbative terms can be treated by stationary pertubation theory. This
technique has been presented in \cite{PhysicaA}. Note that for a simple RWA
two-level system of Rabi frequency $\Omega $ and detuning $\Delta $, the
perturbative regime is such that $\Omega \ll \left| \Delta \right| $ and the
resonant regime such that $\Omega \gtrsim $ $\left| \Delta \right| $. We
classify the different regimes as functions of the ranges of the field
amplitudes and of the detunings. In the following, we have normalized all
the quantities with respect to $\delta .$

\section{Weak field regime}

\label{Weak}

The {\it weak field regime} occurs when $\Omega _{1}\left( t\right) ,\Omega
_{2}\left( t\right) <\delta $. Note that when one has $\Delta _{1}=\Delta
_{2}$ additionally, this regime coincides with a {\it strong spin coupling}
since we have then $4\left| \xi \right| >\Omega _{1}\left( t\right) ,\Omega
_{2}\left( t\right) $. In this case of weak field regime, we can intuitively
analyze the different regimes with respect to the range of the detunings
using the diagram of linkage patterns (Fig. \ref{linkage}). Five relevant
regimes (bounded by dashed lines) have been collected in Fig. \ref{regimes},
depending on the quasi-resonances:

\begin{description}
\item ($A$) The transition 1-2 is quasi-resonant with $\omega _{1}$ and
perturbed by $\omega _{2}$, 2-3 is quasi-resonant with $\omega _{2}$ and
perturbed by $\omega _{1}$;

\item ($B$) 1-2 is quasi-resonant with $\omega _{2}$ and perturbed by $%
\omega _{1}$, 2-3 is quasi-resonant with $\omega _{1}$ and perturbed by $%
\omega _{2}$;

\item ($C$) 1-2 and 2-3 are both quasi-resonant with $\omega _{2}$, and
perturbed by $\omega _{1}$;

\item ($D$) 1-2 is quasi-resonant with $\omega _{2}$ and perturbed by $%
\omega _{1}$, 2-3 is perturbed by $\omega _{2}$ and $\omega _{1}$;

\item ($\tilde{C}$) 1-2 and 2-3 are both quasi-resonant with $\omega _{1}$,
and perturbed by $\omega _{2}$;

\item ($\tilde{D}$) 1-2 is quasi-resonant with $\omega _{1}$ and perturbed
by $\omega _{2}$, 2-3 is perturbed by $\omega _{1}$ and $\omega _{2}$.
\end{description}

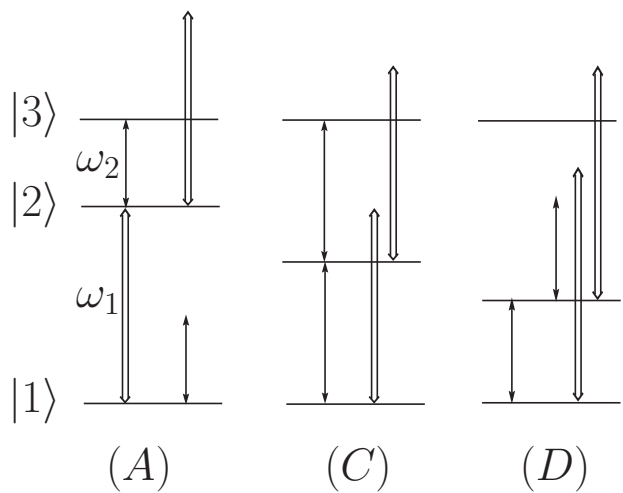
\begin{figure}[tbp]
\caption{Diagram of linkage patterns for the three regimes (in the resonant
case): $A$ ($\Delta _{1}=\Delta _{2}=0$), $C$ ($\Delta _{1}=-\protect\delta %
,\ \Delta _{2}=0$) and $D$ ($\Delta _{1}=-\protect\delta ,\ \Delta _{2}=-%
\protect\delta $).}
\label{linkageACD}
\end{figure}

In the exact resonant cases, we have represented the regimes $A$, $C$ and $D$
in Fig. (\ref{linkageACD}).

As shown schematically in Fig. \ref{regimes}, the above regimes can be
roughly bounded by 
\begin{equation}
\Delta _{1}=\pm \delta /2,\quad \Delta _{1}=-3\delta /2\quad \Delta _{2}=\pm
\delta /2,\quad \Delta _{2}=3\delta /2.
\end{equation}
By the symmetry (\ref{symmetry}), we recover the regime $B$ from the regime $%
A$, $\tilde{C}$ from $C$, $\tilde{D}$ from $D$ (exchanging additionally $%
\Omega _{1}$ and $\Omega _{2}$).

The regimes $A$ and $B$ are STIRAP-like regimes; $D$ and $\tilde{D}$ are
SCRAP-like regimes.

We do not consider other regimes where the state $\left| 1\right\rangle $ is
almost not depopulated by adiabatic passage.

The line $\Delta =-\delta /2$ appears as a dashed line in Figs \ref%
{contours1} and \ref{contours2}, where the restriction $\Delta _{1}=\Delta
_{2}$ has been considered.

\subsection{Regime A}

When the transition 1-2 is quasi-resonant with the frequency $\omega _{1}$
and the transition 2-3 quasi-resonant with the frequency $\omega _{2}$, the
process can be analyzed as the channel A {\it perturbed} (in the sense
non-resonant perturbation theory) by the channel B. We refer to it as the
regime A as shown in Figs \ref{contours1} and \ref{contours2}, where it is
roughly bounded by the dashed lines $\Delta =-\delta /2$, $\Delta =\delta /2$
(not shown) and $\Omega _{0}=\delta $. This regime is approximately
characterized by the following effective Hamiltonian in the basis $\{\left|
1;0,0\right\rangle $,$\left| 2;-1,0\right\rangle $,$\left|
3;-1,-1\right\rangle \}$ \cite{OE99}: 
\begin{equation}
\widetilde{{\sf H}}_{\text{c}}^{\text{A}}=\frac{1}{2}\left[ 
\begin{array}{ccc}
-\frac{\left[ \Omega _{2}\left( t\right) \right] ^{2}}{2\left( \delta
+\Delta _{1}\right) } & \Omega _{1}\left( t\right) & 0 \\ 
\Omega _{1}\left( t\right) & 2\Delta _{1}+\frac{\left[ \Omega _{2}\left(
t\right) \right] ^{2}}{2\left( \delta +\Delta _{1}\right) }+\frac{\left[
\Omega _{1}\left( t\right) \right] ^{2}}{2\left( \delta -\Delta _{2}\right) }
& \Omega _{2}\left( t\right) \\ 
0 & \Omega _{2}\left( t\right) & 2\left( \Delta _{1}+\Delta _{2}\right) -%
\frac{\left[ \Omega _{1}\left( t\right) \right] ^{2}}{2\left( \delta -\Delta
_{2}\right) }%
\end{array}
\right]  \label{A}
\end{equation}
which corresponds to the Hamiltonian characterizing the channel A with
additional time dependent Stark shifts (on the diagonal) induced by the
channel B. Note that this effective Hamiltonian is less precise for bigger $%
\Omega _{1}$ or $\Omega _{2}$ approaching $\delta $.

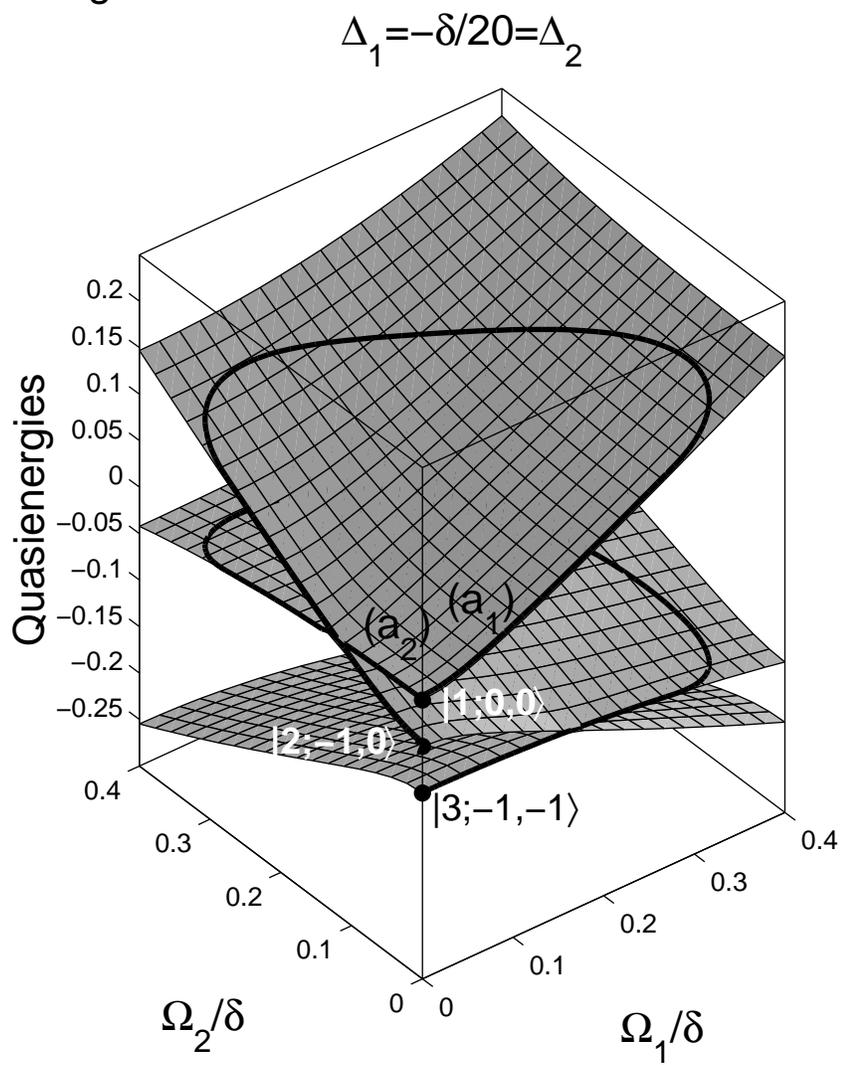
\begin{figure}[tbp]
\caption{Quasienergy surfaces (in units of $\protect\delta $) as functions
of $\Omega _{1}/\protect\delta $ and $\Omega _{2}/\protect\delta $ for $%
\Delta _{1}=\Delta _{2}=-\protect\delta /20$. The path denoted $a_{1}$
(sequence 1), for $\Omega _{0}=0.35\protect\delta $, connect the states $%
|1\rangle $ and $|2\rangle $ with the absorption of one $\protect\omega %
_{1}- $photon. The path denoted $a_{2}$ (sequence 2), for $\Omega _{0}=0.35%
\protect\delta $ connect the states $|1\rangle $ and $|3\rangle $ with the
absorptions of one $\protect\omega _{1}-$photon and of one $\protect\omega %
_{2}-$photon.}
\label{quasiA1}
\end{figure}

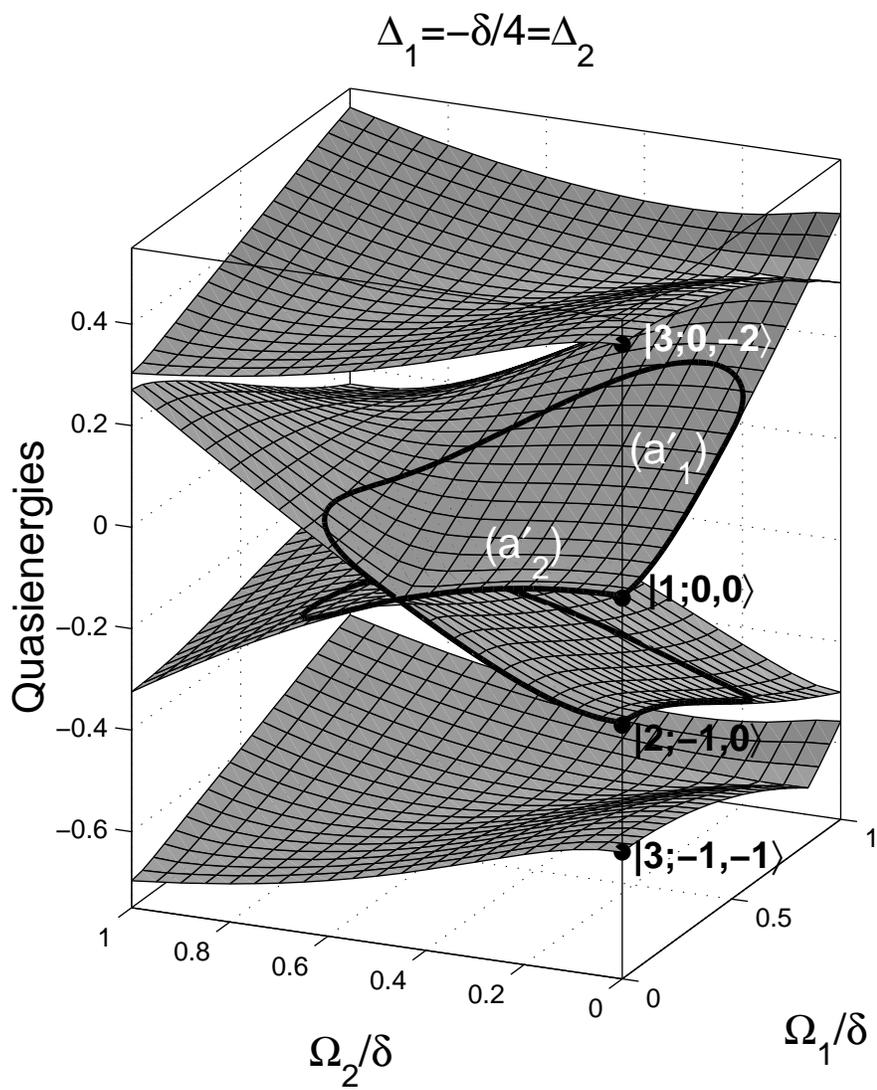
\begin{figure}[tbp]
\caption{Quasienergy surfaces as functions of $\Omega _{1}/\protect\delta $
and $\Omega _{2}/\protect\delta $ for $\Delta _{1}=\Delta _{2}=-\protect%
\delta /4$. The two different paths, denoted $a_{1}$ and $a_{2}$ (for $%
\Omega _{0}=0.7\protect\delta $) depending on the sequence of the pulses
connect the states $|1\rangle $ and $|2\rangle $ with the absorption of one $%
\protect\omega _{1}-$photon.}
\label{quasiA2}
\end{figure}

Before analyzing the dynamics given by this Hamiltonian (\ref{A}), we recall
the results in the limit case of a very weak field $\Omega _{1}\left(
t\right) ,\Omega _{2}\left( t\right) \ll \delta $ obtained in Ref. \cite%
{topology}. In this case the perturbative terms can be neglected and the
Hamiltonian becomes 
\begin{equation}
\widetilde{{\sf H}}_{\text{c}}^{\text{A}}\rightarrow \frac{1}{2}\left[ 
\begin{array}{ccc}
0 & \Omega _{1}\left( t\right) & 0 \\ 
\Omega _{1}\left( t\right) & 2\Delta _{1} & \Omega _{2}\left( t\right) \\ 
0 & \Omega _{2}\left( t\right) & 2\left( \Delta _{1}+\Delta _{2}\right)%
\end{array}
\right] .
\end{equation}
This resulting effective Hamiltonian corresponds to the channel A alone. The
topology of the energy surfaces of this Hamiltonian has been analyzed in %
\cite{topology}. It has been shown that the adiabatic transfer to state $%
\left| 2\right\rangle $ is topologically allowed for 
\[
\Delta _{1}\Delta _{2}>0. 
\]
The topological analysis shows moreover that for the sequence 1 the region
of this process is bounded in the parameter space by the curves 
\[
\Omega _{0}=2\sqrt{\Delta _{1}\left( \Delta _{1}+\Delta _{2}\right) } 
\]
and for the sequence 2 by the curves 
\[
\Omega _{0}=2\sqrt{\Delta _{1}\left( \Delta _{1}+\Delta _{2}\right) }\quad 
\text{and}\quad \Omega _{0}=2\sqrt{\Delta _{2}\left( \Delta _{1}+\Delta
_{2}\right) }. 
\]
Taking now into account the perturbation by the channel B [Hamiltonian (\ref%
{A})] leads to two kinds of topology as shown in Figs \ref{quasiA1} and \ref%
{quasiA2}, where the surfaces of quasienergies as functions of the
normalized Rabi frequencies $\Omega _{1}/\delta $ and $\Omega _{2}/\delta $
respectively for $\Delta =\Delta _{1}=\Delta _{2}=-\delta /20$ and $\Delta
=\Delta _{1}=\Delta _{2}=-\delta /4$ have been displayed. The eigenvalues of
(\ref{A}) (not shown) fit these surfaces well except in Fig. \ref{quasiA2}
when $\Omega _{1}\sim \delta $ and $\Omega _{2}\sim \delta $ because of an
additional dynamical resonance (i.e. a resonance occuring beyond a threshold
of the field amplitudes) \cite{PhysicaA,confusion} which involves the
surface connected with $\left| 3;0,-2\right\rangle $ (which corresponds to
the surface connected to $\left| 3;-1,-1\right\rangle $ and translated of $%
\delta $) and the surface right below.

Fig. \ref{quasiA1} shows that the two conical intersections, one occuring
for $\Omega _{1}=0$, the other one for $\Omega _{2}=0$, bound the adiabatic
connection between the initial state $\left| 1;0,0\right\rangle $ and the
target state $\left| 2;-1,0\right\rangle $. More precisely, for the sequence
1, the conical intersection occuring for $\Omega _{1}=0$ is favorable for
this adiabatic connectivity. The path denoted $a_{1}$ (also corresponding to
the cross $a_{1}$ of Fig. \ref{contours1}) is an example for the complete
transfer. On the other hand, for the sequence 2, the conical intersection
occuring for $\Omega _{1}=0$ is also favorable but the one occuring for $%
\Omega _{2}=0$ is detrimental since it makes $\left| 1;0,0\right\rangle $
connect with $\left| 3;-1,-1\right\rangle $. The path denoted $a_{2}$ is an
example for the complete transfer to $\left| 3;-1,-1\right\rangle $ (also
corresponding to the cross $a_{2}$ of Fig. \ref{contours2}).

For a bigger detuning (in absolute value), the topology is different as
shown in Fig. \ref{quasiA2}. The previous conical intersection occuring for $%
\Omega _{2}=0$ has now disappeared and another one involving the
surfaces connected to $\left| 1;0,0\right\rangle $ and $\left|
3;0,-2\right\rangle $ has appeared.
The two conical intersections, the one occuring for $%
\Omega _{1}=0$ and the other one for $\Omega _{2}=0$, are involved for the
adiabatic connection between the initial state $\left| 1;0,0\right\rangle $
and the target state $\left| 2;-1,0\right\rangle $. More precisely, for the
sequence 1, these two conical intersections bound this adiabatic connection;
the path denoted $a_{1}^{\prime }$ (also corresponding to the cross $%
a_{1}^{\prime }$ of Fig. \ref{contours1}) is an example for the complete
transfer. On the other hand, for the sequence 2 only the conical
intersection occuring for $\Omega _{1}=0$ bounds now the adiabatic
connection; the path denoted $a_{2}^{\prime }$ is an example for the
complete transfer (also corresponding to the cross $a_{2}^{\prime }$ of Fig. %
\ref{contours2}).

Using the effective Hamiltonian (\ref{A}), the position of the previous conical
intersections, for $\Omega _{1}=0$ and $\Omega _{2}=0$ respectively, lead to
the three boundaries for the sequence 1 
\begin{mathletters}
\label{boundA1}
\begin{eqnarray}
\Delta &=&\frac{\Omega _{2}}{16\delta }\left( -5\Omega _{2}\pm \sqrt{9\left(
\Omega _{2}\right) ^{2}+32\delta ^{2}}\right) ,  \label{boundA1left} \\
\Omega _{1} &=&\sqrt{2(\delta -\Delta )\left[ 2\left( \delta +\Delta \right)
-\sqrt{2\left( \delta +\Delta \right) }\right] }.  \label{boundA1right}
\end{eqnarray}
The delay between the pulses has been chosen sufficiently large such that it
is a good approximation to consider that the adiabatic connectivity is quite
well described by the value of the peak amplitudes. Thus we have displayed
these boundaries in Fig. \ref{contours1} as full lines, with $\Omega
_{0}=\Omega _{2}$ for (\ref{boundA1left}) and with $\Omega _{0}=\Omega _{1}$
for (\ref{boundA1right}). They globally bound the lower and upper part of
the island of good transfer of the regime A observed in the numerical
computation. This island is crossed by the line of resonance $\Delta =0$
around which the transfer to $\left| 2\right\rangle $ depends on the pulse
areas, as shown by small oscillating islands.

For the sequence 2, the conical intersections involved give the following
boundaries 
\end{mathletters}
\begin{mathletters}
\label{boundA2}
\begin{eqnarray}
\Delta &=&\frac{\Omega _{2}}{16\delta }\left( -5\Omega _{2}-\sqrt{9\left(
\Omega _{2}\right) ^{2}+32\delta ^{2}}\right) ,  \label{boundA2down} \\
\Omega _{1} &=&\sqrt{(\delta -\Delta )\left[ 4\Delta +\delta \pm \sqrt{%
\delta \left( \delta +8\Delta \right) }\right] },\quad \text{for }\Delta <0.
\label{boundA2up}
\end{eqnarray}
These curves are displayed in fig. \ref{contours2}, with $\Omega _{0}=\Omega
_{2}$ for (\ref{boundA2down}) and with $\Omega _{0}=\Omega _{1}$ for (\ref%
{boundA2up}). They give a good prediction of the island of good transfer of
the regime A observed numerically.

For the two sequences, the islands of good transfer to the state $\left|
2\right\rangle $ of the regime A occur with absorption of one $\omega _{1}-$%
photon.

\subsection{Regime B}

This regime is characterized by the transition 1-2 quasi-resonant with the
frequency $\omega _{2}$ and the transition 2-3 quasi-resonant with the
frequency $\omega _{1}$. This process can be analyzed as the channel B {\it %
perturbed} (in the sense of non-resonant perturbation theory) by the channel
A and is described by the effective Hamiltonian $\widetilde{{\sf H}}_{\text{c%
}}^{\text{B}}={\cal T}\widetilde{{\sf H}}_{\text{c}}^{\text{A}}$: 
\end{mathletters}
\begin{equation}
\widetilde{{\sf H}}_{\text{c}}^{\text{B}}=\frac{1}{2}\left[ 
\begin{array}{ccc}
-\frac{\left[ \Omega _{1}\left( t\right) \right] ^{2}}{2\Delta _{1}} & 
\Omega _{2}\left( t\right) & 0 \\ 
\Omega _{2}\left( t\right) & 2\left( \Delta _{1}+\delta \right) +\frac{\left[
\Omega _{1}\left( t\right) \right] ^{2}}{2\Delta _{1}}-\frac{\left[ \Omega
_{2}\left( t\right) \right] ^{2}}{2\Delta _{2}} & \Omega _{1}\left( t\right)
\\ 
0 & \Omega _{1}\left( t\right) & 2\left( \Delta _{1}+\Delta _{2}\right) +%
\frac{\left[ \Omega _{2}\left( t\right) \right] ^{2}}{2\Delta _{2}}%
\end{array}
\right] .  \label{B}
\end{equation}
The regions of high transfer efficiency to the state $\left| 2\right\rangle $
are bounded in the same manner as in the regime A by the lines (\ref{boundA1}%
) and (\ref{boundA2}) to which we apply the transformation ${\cal T}$ (\ref%
{symmetry}).

\subsection{Regime C}

The regime $C$ is characterized by a mixture of regimes $A$ and $B$ for
which the transitions 1-2 and 2-3 are both quasi-resonant with the same
frequency $\omega _{2}$. As long as the $\omega _{1}-$field is perturbative
for both transitions, we have the following effective Hamiltonian 
\begin{equation}
\widetilde{{\sf H}}_{\text{c}}^{\text{C}}=\frac{1}{2}\left[ 
\begin{array}{ccc}
-\frac{\left[ \Omega _{1}\left( t\right) \right] ^{2}}{2\Delta _{1}} & 
\Omega _{2}\left( t\right) & 0 \\ 
\Omega _{2}\left( t\right) & 2\left( \Delta _{1}+\delta \right) +\frac{\left[
\Omega _{1}\left( t\right) \right] ^{2}}{2\Delta _{1}}-\frac{\left[ \Omega
_{1}\left( t\right) \right] ^{2}}{2(\Delta _{2}-\delta )} & \Omega
_{2}\left( t\right) \\ 
0 & \Omega _{2}\left( t\right) & 2\left( \Delta _{1}+\Delta _{2}+\delta
\right) +\frac{\left[ \Omega _{1}\left( t\right) \right] ^{2}}{2(\Delta
_{2}-\delta )}%
\end{array}
\right] ,  \label{C}
\end{equation}
in the basis $\{\left| 1;0,0\right\rangle $,$\left| 2;0,-1\right\rangle
,\left| 3;0,-2\right\rangle \}$. No efficient transfer is observed in this
regime.

\subsection{Regime D}

This regime is such that the only quasiresonance is between the states 1 and
2 with $\omega _{2}$. In this case, in the basis $\{\left|
1;0,0\right\rangle $,$\left| 2;0,-1\right\rangle $,$\left|
3;0,-2\right\rangle \}$ we can construct an effective Hamiltonian from the
previous one [Eq. (\ref{C})] considering that the $\omega _{2}-$field is
perturbative for the transistion 2-3: 
\begin{equation}
\widetilde{{\sf H}}_{\text{c}}^{\text{D}}=\frac{1}{2}\left[ 
\begin{array}{ccc}
-\frac{\left[ \Omega _{1}\left( t\right) \right] ^{2}}{2\Delta _{1}} & 
\Omega _{2}\left( t\right)  & 0 \\ 
\Omega _{2}\left( t\right)  & 2\left( \Delta _{1}+\delta \right) +\frac{%
\left[ \Omega _{1}\left( t\right) \right] ^{2}}{2\Delta _{1}}-\frac{\left[
\Omega _{1}\left( t\right) \right] ^{2}}{2(\Delta _{2}-\delta )}-\frac{\left[
\Omega _{2}\left( t\right) \right] ^{2}}{2\Delta _{2}} & 0 \\ 
0 & 0 & 2\left( \Delta _{1}+\Delta _{2}+\delta \right) +\frac{\left[ \Omega
_{1}\left( t\right) \right] ^{2}}{2(\Delta _{2}-\delta )}+\frac{\left[
\Omega _{2}\left( t\right) \right] ^{2}}{2\Delta _{2}}%
\end{array}%
\right] .  \label{D}
\end{equation}%
We can remark that this Hamiltonian is valid for the field amplitude $\Omega
_{2}$ below the position of the resonance occuring between the transition
2-3 and the $\omega _{2}-$ field that can be estimated by 
\begin{equation}
\Omega _{2}^{\text{r}}\equiv 2\sqrt{\Delta _{2}\left( \Delta _{1}+\Delta
_{2}+\delta \right) }\quad \text{and\quad }\Delta _{1}+2\Delta _{2}+\delta
\leqslant 0.  \label{threshold2}
\end{equation}%
This limit is represented as the bent dashed line crossing the figure
horizontally in Figs \ref{contours1} and \ref{contours2} (with $\Omega
_{0}=\Omega _{2}^{\text{r}}$). Below this limit, one is allowed to decouple
the states $\left| 2;0,-1\right\rangle $ and $\left| 3;0,-2\right\rangle $
from the Hamiltonian (\ref{C}). A more detailed analysis of this regime
shows that a {\it dynamical resonance} between the transition 1-2 and the $%
\omega _{1}-$field, induced by the $\omega _{2}-$field occurs approximately
for 
\begin{equation}
\Omega _{2}=\Omega _{2}^{\text{dr}}\equiv \sqrt{-\Delta _{1}\left( \Delta
_{1}+2\delta \right) }.  \label{threshold1}
\end{equation}%
It is obtained when the difference of the dressed eigenvalues connected to $%
\left| 1;0,0\right\rangle $ and $\left| 2;0,-1\right\rangle $ (calculated
without the Stark shifts) compensates the difference of the frequencies $%
\delta $. This additional resonance is described as dynamical since it
occurs beyond a threshold of the $\omega _{2}-$field amplitude. It is
represented as the bent dashed line crossing the figure vertically (which
separates the regimes $D$ and $D^{\prime }$) in Figs \ref{contours1} and \ref%
{contours2} with $\Omega _{0}=\Omega _{2}^{\text{dr}}$. The Hamiltonian (\ref%
{D}) is thus approximately valid {\it before} the dynamical resonance (\ref%
{threshold1}).

Below this dynamical resonance, this Hamiltonian (\ref{D}) is very similar
to the one describing the process named SCRAP (acronym for Stark chirped
Rapid adiabatic passage) between the states $\left| 1;0,0\right\rangle $ and 
$\left| 2;0,-1\right\rangle $ \cite{Yatsenko_PRA99}. The pump of this
process is here $\Omega _{2}$ and the Stark pulse $\Omega _{1}$. We have
here $\Omega _{2}$ acting additionally as a Stark pulse.

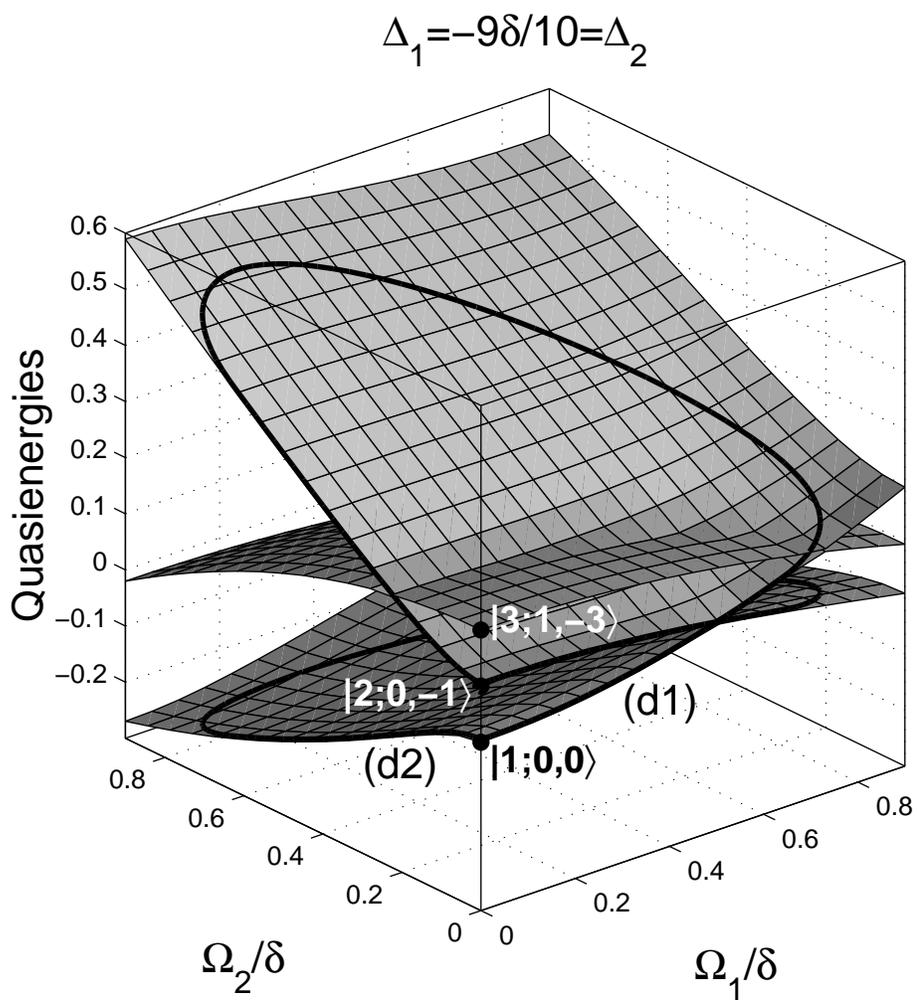
\begin{figure}[tbp]
\caption{Quasienergy surfaces as functions of $\Omega _{1}/\protect\delta $
and $\Omega _{2}/\protect\delta $ for $\Delta _{1}=\Delta _{2}=-9\protect%
\delta /10$. Two different paths (denoted $d_{1}$ and $d_{2}$) for $\Omega
_{0}=0.8\protect\delta $ connect the states $|1\rangle $ and $|2\rangle $
with the absorption of one $\protect\omega _{2}-$photon.}
\label{quasiD}
\end{figure}

It is important to note that when $\Delta _{1}=-\delta $, the field $\omega
_{2}$ is exactly in resonance with the transition 1-2, and it cannot induce
any complete population transfer from $\left| 1\right\rangle $ to $\left|
2\right\rangle $. Below this boundary (plotted as a full line in Figs \ref%
{contours1} and \ref{contours2}), i.e. for $\Delta _{1}<-\delta $, the
topology does not allow the transfer from $\left| 1\right\rangle $ to $%
\left| 2\right\rangle $. Above this aboudary ($\Delta _{1}>-\delta $), the
transfer is possible as shown by the surfaces of quasienergies (for $\Delta
=\Delta _{1}=\Delta _{2}=9\delta /10)$ in Fig. \ref{quasiD}. The eigenvalues
of (\ref{D}) (not shown) fit well these surfaces below the dynamical
resonances $\Omega _{2}<\Omega _{2}^{r}.$ Fig. \ref{quasiD} shows that the
conical intersection for $\Omega _{2}=0$ between the surfaces connected to $%
\left| 1;0,0\right\rangle $ and the target state $\left| 2;0,-1\right\rangle 
$ bound the adiabatic connection between these states. This characterizes a
transfer to the state $\left| 2\right\rangle $ with absorption of one $%
\omega _{2}-$photon. This boundary is calculated from the effective
Hamiltonian (\ref{D}): 
\[
\Omega _{1}=2\sqrt{\Delta \frac{\Delta ^{2}-\delta ^{2}}{2\delta -\Delta }}. 
\]
It is plotted in Figs \ref{contours1} and \ref{contours2} as a full line in
the region $D$ and bounds the upper island of good transfer of this region.

The cases beyond the dynamical resonance are studied in the next section.

\section{Strong field regimes}

\label{Strong}

The {\it strong field regime} occurs when $\Omega _{1}\left( t\right)
,\Omega _{2}\left( t\right) \gtrsim \delta $. For $\Delta _{1}=\Delta _{2}$,
this corresponds to a {\it weak spin coupling} since one has then $4\left|
\xi \right| \lesssim \Omega _{1}\left( t\right) ,\Omega _{2}\left( t\right) $%
. More resonances occur in this case and the previous effective Hamiltonians
are no more valid. We will study in detail the interesting regime $D^{\prime
}$ which gives quite large areas of transfer to state $\left| 2\right\rangle 
$.

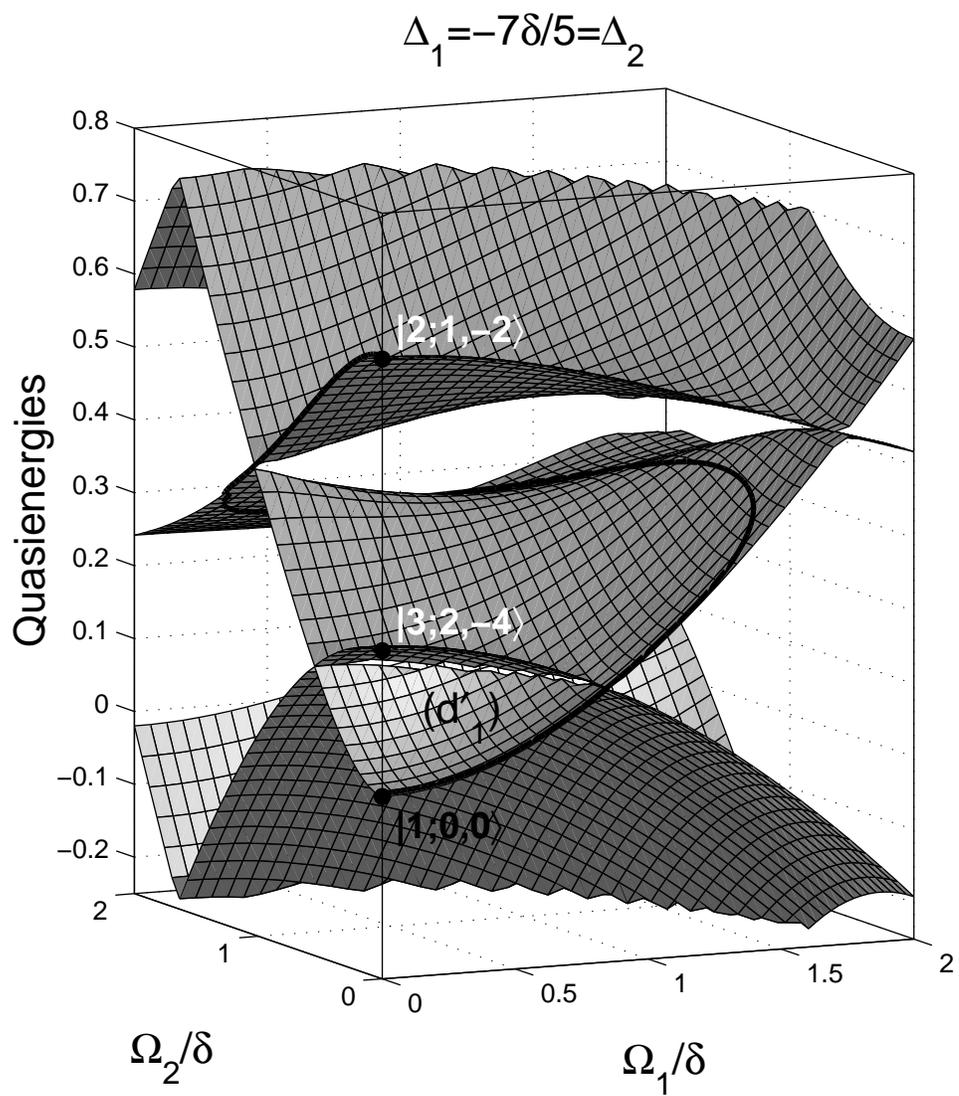
\begin{figure}[tbp]
\caption{Quasienergy surfaces as functions of $\Omega _{1}/\protect\delta $
and $\Omega _{2}/\protect\delta $ for $\Delta _{1}=\Delta _{2}=-7\protect%
\delta /5$. Two different paths (denoted $d_{1}^{\prime }$ and $%
d_{2}^{\prime }$) for $\Omega _{0}=3\protect\delta /2$ connect the states $%
|1\rangle $ and $|2\rangle $ with the absorption of two $\protect\omega %
_{2}- $photon and the emission of one $\protect\omega _{1}-$photon.}
\label{quasiDp}
\end{figure}

This regime is located below the resonance (\ref{threshold2}) and beyond the
dynamical resonance (\ref{threshold1}), when the transition 1-2 is
quasi-resonant with both the $\omega _{1}-$ and $\omega _{2}-$fields and
when the transition 2-3 is not resonant with either the $\omega _{1}-$field
or the $\omega _{2}-$field. This regime is thus characterized by the
effective dressed Hamiltonian 
\begin{equation}
{\sf K}^{\text{D'}}=-\text{i}\delta \frac{\partial }{\partial \theta }+\frac{%
1}{2}\left[ 
\begin{array}{ccc}
0 & \Omega _{2}\left( t\right) +e^{\text{i}\theta }\Omega _{1}\left(
t\right)  & 0 \\ 
\Omega _{2}\left( t\right) +e^{-\text{i}\theta }\Omega _{1}\left( t\right) 
& 2\left( \Delta _{1}+\delta \right) -\frac{\left[ \Omega _{1}\left(
t\right) \right] ^{2}}{2(\Delta _{2}-\delta )}-\frac{\left[ \Omega
_{2}\left( t\right) \right] ^{2}}{2\Delta _{2}} & 0 \\ 
0 & 0 & 2\left( \Delta _{1}+\Delta _{2}+\delta \right) +\frac{\left[ \Omega
_{1}\left( t\right) \right] ^{2}}{2(\Delta _{2}-\delta )}+\frac{\left[
\Omega _{2}\left( t\right) \right] ^{2}}{2\Delta _{2}}%
\end{array}%
\right] .  \label{Dp}
\end{equation}%
It is equivalent to a two-level system driven by a bichromatic field \cite%
{PRA_RC} with additional Stark shifts. The surfaces of quasienergies as
functions of the normalized Rabi frequencies $\Omega _{1}/\delta $ and $%
\Omega _{2}/\delta $ (for $\Delta =\Delta _{1}=\Delta _{2}=-7\delta /5$) are
displayed in Fig. \ref{quasiDp}. This figure shows that the two conical
intersections, one for $\Omega _{1}=0$ and one for $\Omega _{2}=0$, bound
the adiabatic connection between the initial state $\left|
1;0,0\right\rangle $ and the target state $\left| 2;1,-2\right\rangle $. We
calculate the boundaries using the effective Hamiltonian (\ref{Dp}), which
are plotted as full lines in Figs \ref{contours1} and \ref{contours2}: 
\begin{eqnarray*}
\Omega _{1} &=&2\sqrt{(\Delta -\delta )\left[ 2\delta -\Delta -2\sqrt{\delta
\left( 2\delta -\Delta \right) }\right] }, \\
\Omega _{2} &=&2\sqrt{\Delta \left[ \delta -\Delta -\sqrt{\delta \left(
\delta -4\Delta \right) }\right] }.
\end{eqnarray*}%
This process corresponds to a multiphoton transfer to the state $\left|
2\right\rangle $, with absorption of two $\omega _{2}-$photons and emission
of one $\omega _{1}-$photon.

The analysis of the topology allows to improve the transfer efficiency. It
shows indeed that a $\omega _{1}-$field amplitude larger than the $\omega
_{2}-$field amplitude is better in this regime since the conical
intersection for $\Omega _{1}=0$ occurs for a larger value than the one for $%
\Omega _{2}=0$.

This process of a two-level system driven by a bichromatic field studied in
Ref \cite{PRA_RC} shows that the transfer can still occur for a stronger
field (i.e. for a weaker spin coupling), but with absorption of more than
two $\omega _{2}-$photons and emission of more than one $\omega _{1}-$%
photon. This result is shown in Fig. \ref{verystrong} where strong field
white islands can be observed. The lower white islands correspond to good
population transfer to the entangled state $\left| 2;k-1,-k\right\rangle $,
with $k=1,2,3,4$ from left to right.

\section{Conclusion}

We have shown that a system of two interacting spins in an external
bichromatic field is equivalent to a three level problem. We have determined
the choices of laser pulses which can give a maximal final population in the
entangled state. The proposed strategies are robust with respect to the
external parameters. We have found that in the parameter space it is
possible to find large islands where the quantum system can be transferred
to the entangled state with a high efficiency. These islands of good
transfer have been characterized by the topology of the surfaces of dressed
states as functions of the parameters.

The methods employed here are quite general and can be applied for a large
variety of systems. We anticipate interesting applications of this method in
quantum computing and quantum communication.

\section*{Acknowledgments}

We acknowledge supports by INTAS 99-00019 and the Conseil R\'{e}gional de
Bourgogne. S.G. thanks support from a CNRS project ''jeunes chercheurs''. RU
thanks support by the Alexander Von Humboldt Foundation and l'universit\'{e}
de Bourgogne for kind hospitality. The authors thank N. Vitanov for usefull
discussions.

%
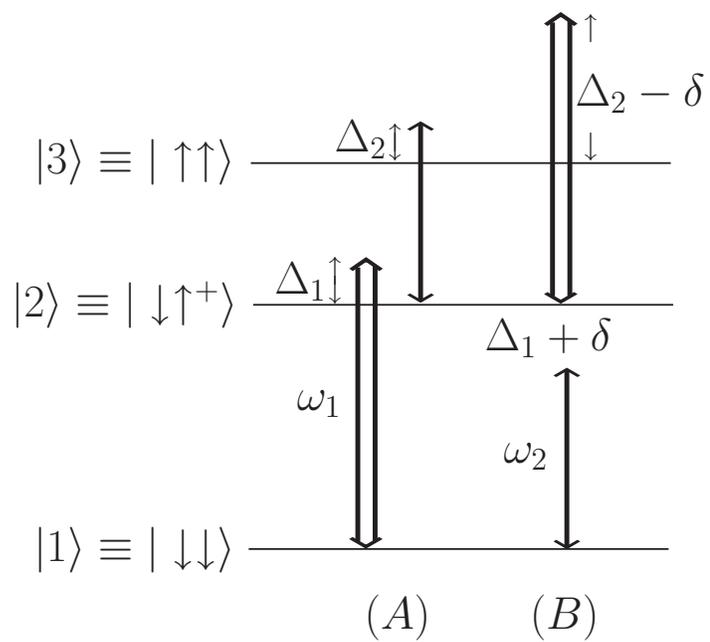
\begin{figure}[tbp]
\caption{Diagram of linkage patterns between the three states showing the
different couplings. Note that $\Delta _{1}$ and $\Delta _{2}$ have been
chosen here negative.}
\label{linkage}
\end{figure}
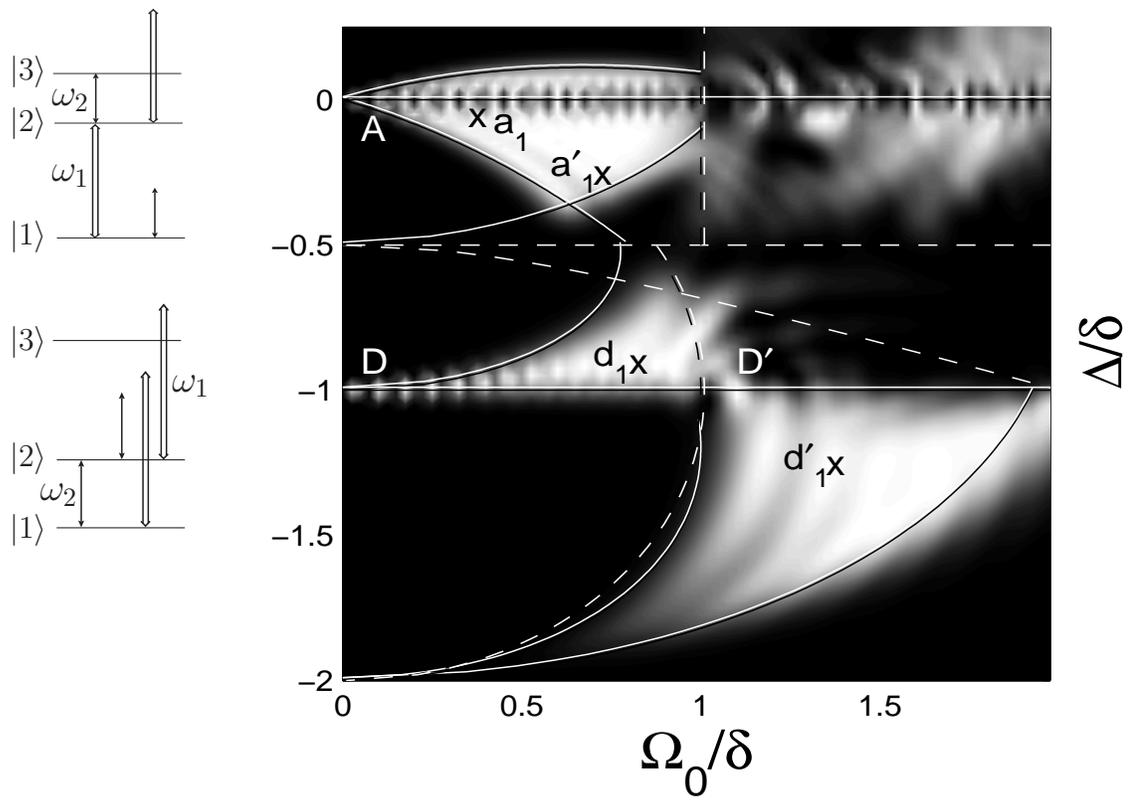
\begin{figure}[tbp]
\caption{Contour map of population transfer efficiency $P_{2}(\infty )$ for
varying peak Rabi frequency $\Omega_{0}/\protect\delta $ and varying
detuning $\Delta /\protect\delta $ (with $\Delta =\Delta _{1}=\Delta _{2}$)
for the sequence 1. White areas correspond to high efficiency transfer
(close to 1) to the entangled state. Dark areas correspond to low efficiency
transfer (close to 0) to the entangled state $|2\rangle $. The dashed lines
separate different regions labelled A, D and D', associated to different
effective Hamiltonians constructed in Sections \ref{Weak} and \ref{Strong}.
The regimes of good population transfer are bounded by full lines predicted
from the topological analysis. The crosses labelled ($a_{1}$), ($%
a_{1}^{\prime }$), ($d_{1}$), and ($d_{1}^{\prime }$) refer to parameters
leading to high efficiency. They also refer to the pathways shown
respectively in Figs \ref{quasiA1}, \ref{quasiA2}, \ref{quasiD} and \ref%
{quasiDp}. The regions A and D}
\label{contours1}
\end{figure}
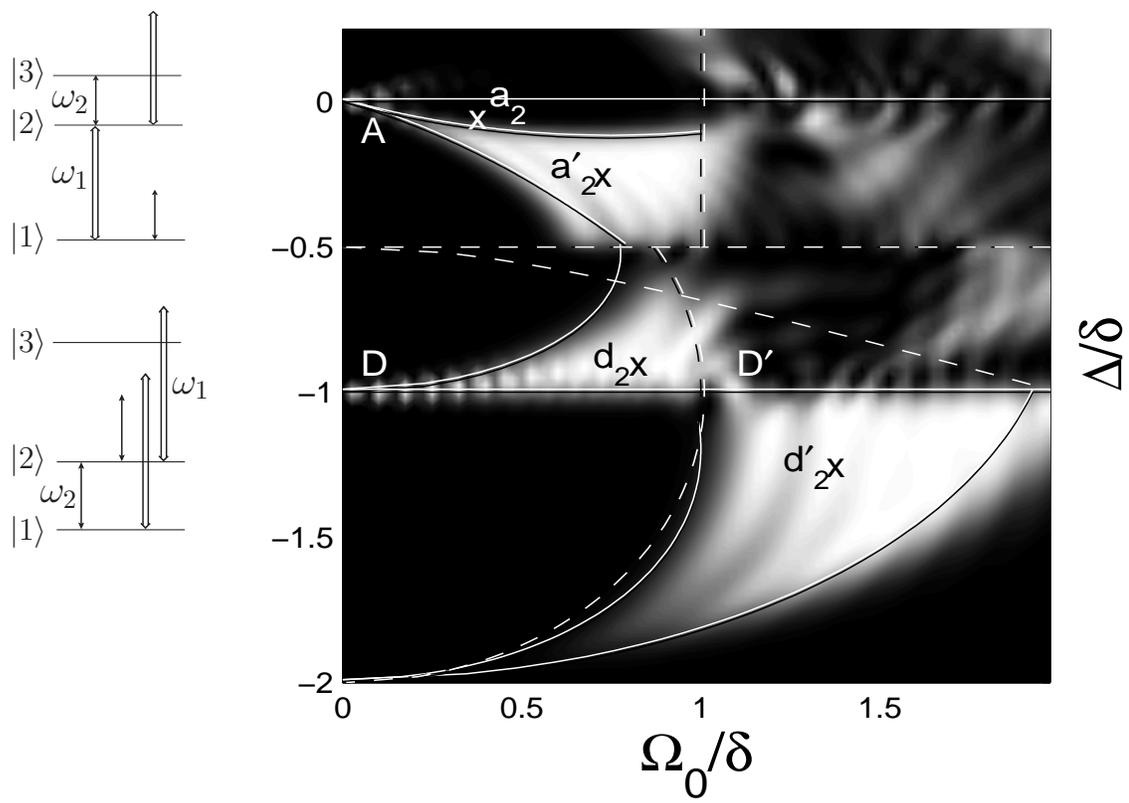
\begin{figure}[tbp]
\caption{Contour map of population transfer efficiency $P_{2}(\infty )$ for
varying peak Rabi frequency $\Omega _{0}/\protect\delta $ and varying
detuning $\Delta /\protect\delta $ for the sequence 2. The cross labelled ($%
a_{2}^{\prime }$) in one region of low efficiency and the ones labelled ($%
a_{2}$), ($c_{2}$) and ($d_{2}$) in the regions of high efficiency refer to
the pathways shown respectively in Figs \ref{quasiA2}, \ref{quasiA1}, \ref%
{quasiD} and \ref{quasiDp}.}
\label{contours2}
\end{figure}
%
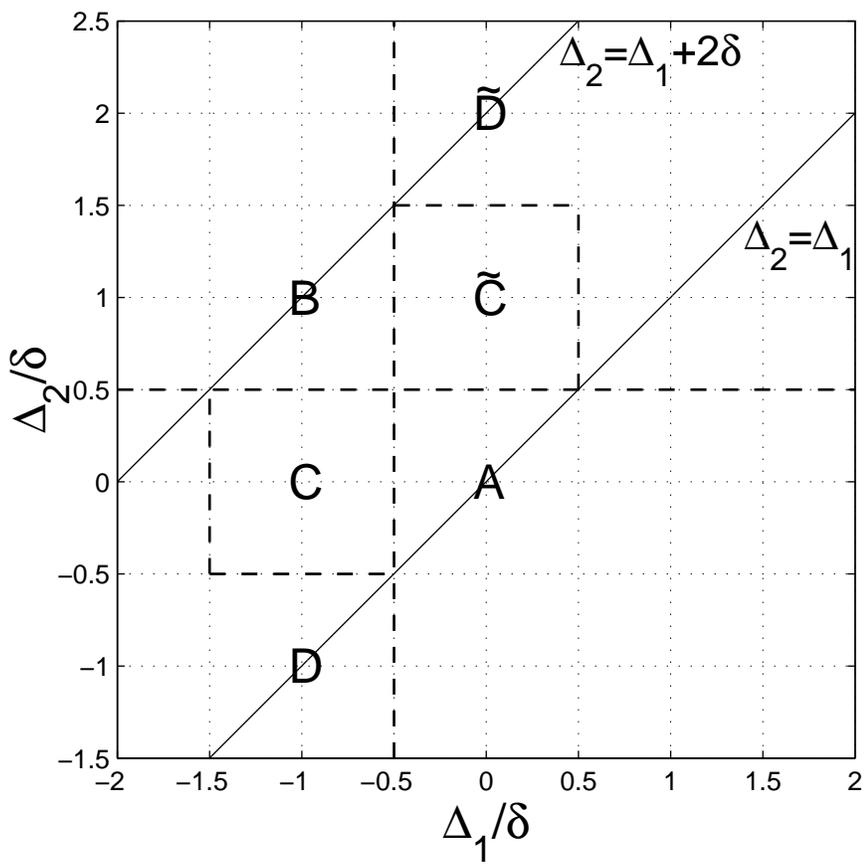
\begin{figure}[tbp]
\caption{Schematic diagram of the regimes for a weak field regime as a
function of the normalized detunings $\Delta _{1}/\protect\delta $ and $%
\Delta _{2}\protect\delta $. The restriction $\Delta _{2}=\Delta _{1}$ has
been used for Figs. \ref{contours1} and \ref{contours2}.}
\label{regimes}
\end{figure}

%

%
%
%
%
%
\begin{figure}[tbp]
\caption{Contour map of population transfer efficiency $P_{2}(\infty )$ as
in Fig. \ref{contours1}, but for stronger field amplitudes.}
\label{verystrong}
\end{figure}

\end{document}